\documentclass[11pt, a4paper]{article}
\usepackage{jheppub}
\usepackage{amsmath}
\usepackage{graphicx}
\usepackage[config=altsf]{subfig}
\usepackage{slashed}
\usepackage{afterpage}
\usepackage{comment}
\usepackage{appendix}
\usepackage{multirow}
\usepackage{dcolumn}
\usepackage{slashed}
\usepackage{soul}
\usepackage{cleveref}
\crefname{figure}{Fig.}{Figs.} 

\usepackage{amsfonts}
\usepackage{amsmath,amssymb}
\usepackage{mathrsfs}
\usepackage{epsfig}
\usepackage{graphicx}
\usepackage{wrapfig}   

\usepackage{url}

\usepackage{hyperref}

\usepackage{enumitem}
\usepackage{ctable}

\usepackage{pgfkeys, pgfsys, pgfcalendar}

\vspace*{3cm}

\title{\bf Helicity amplitude for the process $q\Bar{q} \rightarrow q\Bar{q} g$ }

\author[a,b]{Vibhu Pandya,}
\author[c]{ Radhika Vinze
}
\author[a]{and Anuradha Misra}

\affiliation[a]{UM-DAE Centre for Excellence in Basic Sciences (CEBS), Vidyanagari, Mumbai-400098, India}
\affiliation[b]{International Centre for Theoretical Sciences (ICTS-TIFR), Tata Institute of Fundamental Research, Shivakote, Hesaraghatta, Bangalore 560089, India.}
\affiliation[c]{Department of Physical Sciences, Indian Institute of Science Education and Research Mohali,
	Knowledge city, Sector 81, Manauli, PO, Sahibzada Ajit Singh Nagar, Punjab, India, 140306}
\emailAdd{anuradha.misra@cbs.ac.in}
\emailAdd{vibhu.pandya@cbs.ac.in}
\emailAdd{radhikavinze@iisermohali.ac.in}

\abstract{ Precision calculations in hadronic processes at high energy colliders are crucial for improving the understanding of the standard phenomena as well as for the discovery of new physics. Spinor-helicity formalism serves as one of the most efficient ways to simplify the calculations of $S$ matrix elements. In this article, we compute the $S$ matrix elements for the process $q\Bar{q}\rightarrow q\Bar{q}g$ mediated by photon and gluon. Ignoring the contribution of $Z$ boson exchange, we show that the calculation of $S$ matrix elements for this process simplifies to a great extent by using spinor-helicity formalism. }

\begin{document}
	\maketitle
	
	\flushbottom
	
	\newpage
	
	\section{Introduction}
	\label{sec:intro}
	In high-energy experiments, when two particles collide, there is a plethora of particles that can be produced in the collision. The scattering amplitudes lead to the most probable outcome of such scattering experiment, and hence connect theoretical predictions of the observables to the experimental measurements. Scattering amplitudes provide the probability of the outcome of a collision of two particles. At hadron colliders, due to the composite nature of initial-state particles, the outcome of the scattering has many possibilities, resulting in a large background in the investigation of an observable in a specific process. Hence, achieving an accurate value of an observable i.e. precision in calculations necessitates managing the background effects. Furthermore one needs to go beyond leading order effects to achieve better accuracy in calculating an observable.
	
	In effort of incorporating the higher order effects, loop diagrams containing virtual particles need to be calculated. However, it makes the structure of scattering amplitudes more complex. Real emission of gluon radiation from the initial or final states in the process also gives rise to a larger number of diagrams. As a result, the number of diagonal as well as interference terms increases, making it difficult to compute the amplitude square using the traditional trace method. To address this issue, one can shift to the spinor-helicity formalism introduced in \cite{DeCausmaecker:1981jtq,Gastmans:1990xh,Xu:1986xb,DeCausmaecker:1981wzb,Berends:1981rb,Kleiss:1985yh}. Calculating scattering amplitudes using the spinor-helicity formalism with its algebraic simplifications reduces the number of terms in the scattering amplitude. The wave functions of fermions in the helicity basis are Dirac spinors in Weyl representation.  In spinor-helicity formalism, the amplitude for the scattering process is written taking into account the specific helicity of the particles. At very high energies, the particles in the initial and final state become effectively massless and helicity basis is found to be suitable for calculating scattering amplitudes. Since the chirality and helicity of the particle at very high energies coincide for massless fermions, most of the matrix elements in the scattering amplitude vanish at the tree level as well as at loop level\cite{Dixon:2013uaa}. Thus, calculation of scattering amplitude gets simplified to a great extent in helicity basis. Hence, for many complex scattering processes at very high energies, use of helicity basis for calculating amplitude leads to considerable simplifications. 

    Helicity amplitude formalism has been implemented to calculate scattering amplitudes of various processes in Quantum Electrodynamics as well as Quantum Chromodynamics. Helicity amplitudes can be automatically obtained in Feynman-diagram (FD) gauge at tree level for an arbitrary gauge model with {MadGraph5\_aMC@NLO}\cite{Hagiwara:2024xdh}. Tree level QCD amplitudes with multi-gluon and a pair of massive fermions are calculated in \cite{Huang:2012gs, Ochirov:2018uyq,Gunion:1985vca}. This formalism is beneficial for calculating higher-order corrections. One-loop contributions are calculated using the spinor-helicity mechanism in \cite{Bern:1993mq,Ferdyan:2024kmx}, while multiloop amplitudes are calculated in QCD using spinor-helicity formalism. NNLO amplitudes are calculated in \cite{Badger:2017jhb,Abreu:2018aqd,Badger:2024sqv,Abreu:2020cwb} and $\text{N}^3$LO amplitudes are calculated in \cite{Caola:2021rqz,Guan:2024hlf,Gehrmann:2023jyv}. Two-loop amplitudes in standard electroweak theory can also be calculated using HELAC code\cite{Kanaki:2000ey}. In case of top quark, helicity amplitudes for tree leel single top quark production are presented in \cite{vanderHeide:2000fx,Campbell:2023fjg} and higher order helicity amplitudes are presented in \cite{Badger:2011yu,Badger:2021owl}.

	
	In this article we consider the real emission of a gluon in the quark-antiquark scattering process i.e. $q\overline{q}\rightarrow q\overline{q}g$.  We neglect the $Z$ boson exchange contribution and calculate the amplitude square for $q\overline{q}\rightarrow q\overline{q}g$ mediated by the photon/gluon in the helicity spinor formalism. We focus on a specific helicity combination, since other helicity combinations can be obtained through parity transformation. It is also assumed here that the initial pair and final pair correspond to different quark flavours, so that only s-channel processes contribute. The use of spinor-helicity formalism makes the calculation of amplitude simpler due to cancellations between some diagrams in the amplitude. The article is planned as follows - In section \ref{intro-process} we present the diagrams for the process $q\overline{q}\rightarrow q\overline{q}g$ in spinor-helicity formalism and set the notations. In section \ref{amp-terms} we calculate the diagrams that are mediated by photon and gluon in spinor-helicity formalism. After using helicity spinor formalism techniques, contributions from some of the diagrams vanish. Considering the contribution from remaining diagrams, we calculate amplitude square for $q\overline{q}\rightarrow q\overline{q} g $ with photon and gluon as mediators in Section \ref{calc_check}.

	\section{Amplitude for $q\overline{q} \rightarrow q\overline{q} g $ in helicity basis}
	\label{intro-process}
\begin{figure}
		\centering
		{\includegraphics[width = 1.75in]{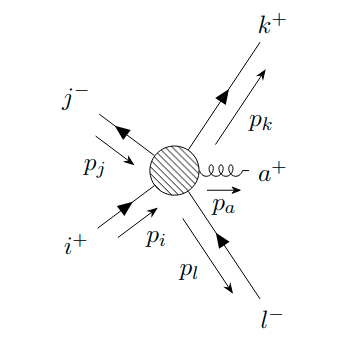}} 
	\caption{quark-anti-quark scattering representation with helicity-spinors}
	\label{fig:diag}
\end{figure}

	We consider a process in which a quark - anti-quark pair produces a pair of quark - anti-quark pair with emission of a real gluon either in the initial state or in the final state. The real gluon emission in $q\overline{q}\rightarrow q\overline{q}$ results in an additional gluon jet at tree level. This process can be mediated by a photon, a gluon, and a Z boson. We focus on the contributions from the photon and gluon mediated diagrams in the amplitude. In order to simplify numerator algebra and calculation of amplitude square, we use the spinor-helicity formalism. We calculate the amplitude square in high-energy limit in which the quark masses can be neglected. Incoming and outgoing particles are represented by helicity eigen-spinors in which $+$ or $-$ indicate the helicities of the particles. We use convention in Fig.~\ref{fig:diag}, where $i, j, k, l$ denote the momenta of external quarks and $a$ denotes momentum of the emitted gluon. We focus on a specific helicity configuration of the external particles as shown in Fig.~\ref{fig:diag}. 

\par \textcolor{red}{}

In order to obtain results for other helicity configurations, one can use parity and charge conjugation symmetry. Parity operator flips all helicities leaving $|\mathcal{M|}^2$ unchanged. Hence, using parity transformation, one can obtain the amplitude for another combination of helicities for example $\mathcal{M}(i^{-}j^{+}k^{-}l^{+}a^{-})$ can be obtained from the amplitudes for diagrams in Fig.\ref{fig:diag} by a parity transformation. Adding contributions from both photon and gluon exchange diagrams, we obtain the total amplitude for the process $q\overline{q}\rightarrow q\overline{q}g$ as
\begin{eqnarray}
	\mathcal{M} = \mathcal{M}_{\gamma} + \mathcal{M}_{g}
	\label{amp}
\end{eqnarray}
neglecting $Z$ boson exchange diagrams. The tree level diagrams are presented in Fig.~\ref{fig:photon-diags} and Fig.~\ref{fig:gluon-diags}.

Each  term in \(\mathcal{M}\) comprasises of a colour factor $\mathcal{C}_{ij}$ and helicity-spinor products $\mathcal{A}_{ij}$
$$\left|\mathcal{M}_{ij}\right|^{2}=\mathcal{A}_{ij}\,\mathcal{C}_{ij}. $$

\section{Helicity amplitude for photon and gluon exchange diagrams}
\label{amp-terms}
\subsection{Photon-Mediated Diagrams}
\label{photon-contrib}

\begin{figure}[t!]
	\centering
	{\includegraphics[width = 1.8in]{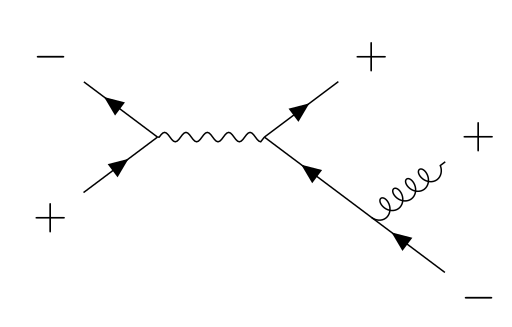}} 
	\hspace{0.7cm}
	{\includegraphics[width = 1.8in]{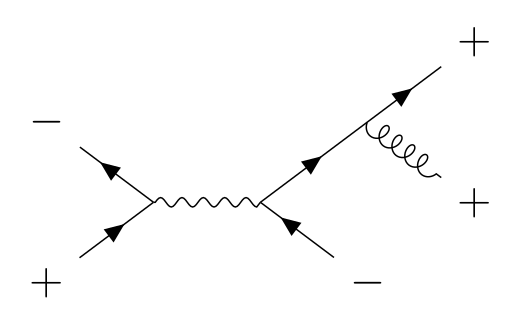}}
	\hspace{0.7cm}
	{\includegraphics[width = 1.8in]{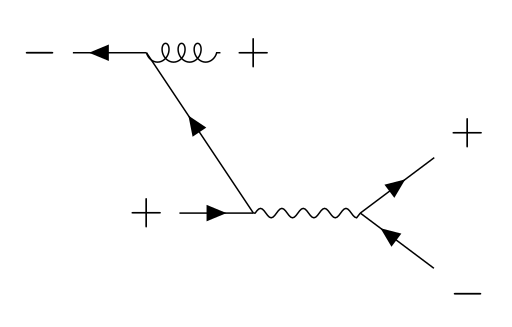}}
	\hspace{0.7cm}
	{\includegraphics[width = 1.8in]{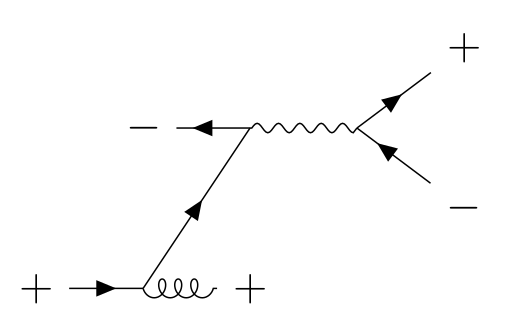}}
	\caption{Photon mediated diagrams for $q\overline{q}\rightarrow q\overline{q}g$ process }
	\label{fig:photon-diags}
\end{figure}

Feynman diagrams in helicity basis for $q\overline{q}\rightarrow q\overline{q}g$ via photon exchange are presented in Fig.~\ref{fig:photon-diags}. Using the Feynman rules
from \cite{Peskin:1995ev}, we write the amplitude for photon mediated diagrams for the process $q\overline{q}\rightarrow q\overline{q}g$ as

\begin{eqnarray}
	\mathcal{M}_\gamma&=&\frac{-iQ_{i}Q_{k}e^{2}g\,[j\,\gamma^{\mu}\,i\rangle\,[k\,\gamma_{\mu}\,( \slashed{a}+\slashed{l_{}}) \slashed{\varepsilon_{+}}\,l\rangle\,T^{a}_{kl}\delta_{ij}}{s_{ij}s_{al}} 
	+\frac{-iQ_{i}Q_{k}e^{2}g\,[j\,\gamma^{\mu}\,i\rangle\,[k\,\slashed{\varepsilon_{+}}\,( \slashed{a}+\slashed{k}) \gamma_{\mu}\,l\rangle\,T^{a}_{kl}\delta_{ij}}{s_{ij}s_{ak}} \nonumber \\
	&~&  +\frac{iQ_{i}Q_{k}e^{2}g\,[k\,\gamma^{\mu}\,l\rangle\,[j\,\slashed{\varepsilon_{+}}\,( \slashed{j_{}}-\slashed{a}) \,\gamma_{\mu}\,i\rangle\,T^{a}_{ji}\delta_{kl}}{s_{kl}s_{aj}}
	+\frac{iQ_{i}Q_{k}e^{2}g\,[k\,\gamma^{\mu}\,l\rangle\,[j\,\gamma_{\mu}\,( \slashed{i_{}}-\slashed{a}) \,\slashed{\varepsilon_{+}}\,i\rangle\,T^{a}_{ji}\delta_{kl}}{s_{kl}s_{ai}} \nonumber\\
	&=&\frac{-\sqrt{2}\,iQ_{i}Q_{k}e^{2}g\,[jk]\langle ql\rangle [la]\langle li\rangle\,T^{a}_{kl}\delta_{ij}}{s_{ij}s_{al}\langle qa\rangle }
	+\frac{\sqrt{2}\,iQ_{i}Q_{k}e^{2}g\,\langle li \rangle [ka] [j\,(\slashed{i_{}}-\slashed{l_{}})\,q\rangle \,T^{a}_{kl}\delta_{ij}}{s_{ij}s_{ak}\langle qa\rangle } \nonumber\\
	&~&  +\frac{\sqrt{2}\,iQ_{i}Q_{k}e^{2}g\,\langle il\rangle [ja] [k\,(\slashed{a}-\slashed{j_{}})q\rangle T^{a}_{ji}\delta_{kl}}{s_{kl}s_{aj}\langle qa\rangle}
	-\frac{\sqrt{2}\,iQ_{i}Q_{k}e^{2}g\,\langle qi\rangle [kj] [a\slashed{i_{}}l\rangle T^{a}_{ji}\delta_{kl}}{s_{kl}s_{ai}\langle qa\rangle} 
	\label{photon-amp}
\end{eqnarray}


Here, $Q_{i}$ and $Q_{k}$ are the electric charges of the initial and final quarks respectively and $Q_{e}=-1$ is the charge of electron.$T_{ij}$'s are QCD generators. As $p_{i}^{2}=p_{j}^{2}=0$ for massless quarks, we have $(p_{i}+p_{j})^{2}=2\,p_{i}\cdot p_{j}=s_{ij}\;\;\; \text{and}\;\;\;(p_{a}-p_{i})^{2}=-2\,p_{a}\cdot p_{i}=-s_{ai} $~. The polarisation vector corresponding to the emitted gluon is given by:
$$ \slashed{\varepsilon}_{+}(k,q)=\frac{|q\rangle[k|\,+\,|k]\langle q|}{\sqrt{2}\langle qk\rangle} $$ 
where $q$ is a reference momentum. Next, we compute the amplitude for the gluon mediated diagrams.

\subsection{Gluon-Mediated Diagrams}
\label{gluon-contrib}

\begin{figure}[t!]
	\centering
	{\includegraphics[width = 1.75in]{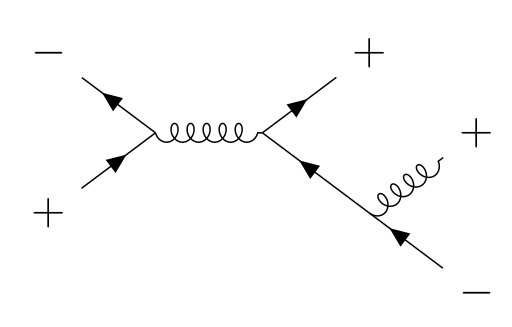}} 
	\hspace{0.6cm}
	{\includegraphics[width = 1.75in]{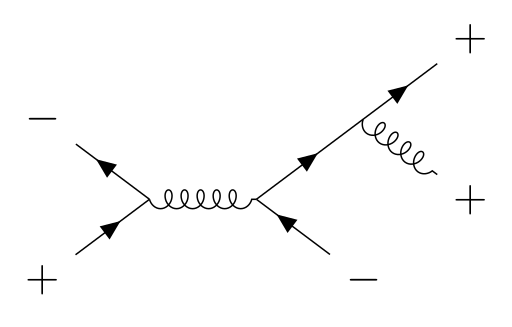}}
	\hspace{0.6cm}
	{\includegraphics[width = 1.75in]{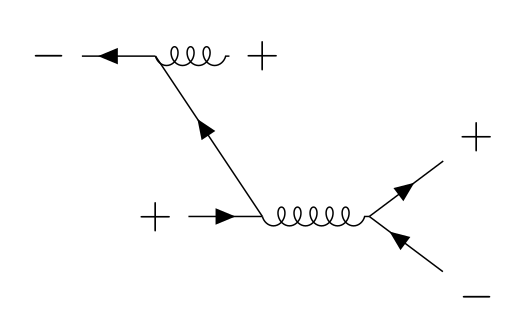}}
	\hspace{0.6cm}
	{\includegraphics[width = 1.75in]{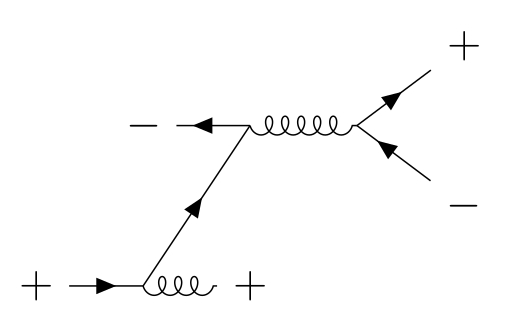}}
	\hspace{0.6cm}
	{\includegraphics[width = 1.75in]{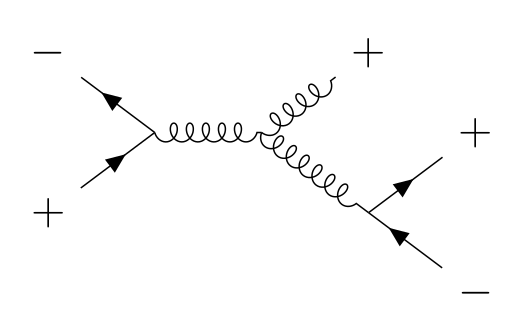}}
	\caption{Gluon mediated diagrams for $q\overline{q}\rightarrow q\overline{q}g$ process }
	\label{fig:gluon-diags}
\end{figure}

The gluon-mediated diagrams that contribute to $q\overline{q}\rightarrow q\overline{q}g$ are shown in Fig.~\ref{fig:gluon-diags}. Using the Feynman rules
from \cite{Dixon:2013uaa, Peskin:1995ev} we write the amplitude for gluon mediated diagrams for the process $q\overline{q}\rightarrow q\overline{q}g$ as

\begin{eqnarray}
	\mathcal{M}_g &=&=\frac{-ig^{3}\,[j\,\gamma^{\mu}\,i\rangle\,[k\,\gamma_{\mu}\,( \slashed{a}+\slashed{l_{}}) \slashed{\varepsilon_{+}}\,l\rangle\,T^{b}_{ji}T^{b}_{kn}T^{a}_{nl}}{s_{ij}s_{al}} +\frac{-ig^{3}\,[j\,\gamma^{\mu}\,i\rangle\,[k\,\slashed{\varepsilon_{+}}\,( \slashed{a}+\slashed{k})\gamma_{\mu}\,l\rangle\,T^{b}_{ji}T^{b}_{nl}T^{a}_{kn}}{s_{ij}s_{ak}} \nonumber \\
	&~& +\frac{ig^{3}\,[k\,\gamma^{\mu}\,l\rangle\,[j\,\slashed{\varepsilon_{+}}\,( \slashed{j_{}}-\slashed{a}) \,\gamma_{\mu}\,i\rangle\,T^{b}_{ni}T^{b}_{kl}T^{a}_{jn}}{s_{kl}s_{aj}} +\frac{ig^{3}\,[k\,\gamma^{\mu}\,l\rangle\,[j\,\gamma_{\mu}\,(\slashed{i_{}}-\slashed{a}) \,\slashed{\varepsilon_{+}}\,i\rangle\,T^{b}_{jn}T^{b}_{kl}T^{a}_{ni}}{s_{kl}s_{ai}} \nonumber\\
	&~& + \frac{g^{3}\,[j\,\gamma^{\mu}\,i\rangle \,[k\,\gamma^{\nu}\,l\rangle\,\varepsilon_{+}^{\rho}\,f^{abc}T^{b}_{ji}T^{c}_{kl}\,\mathcal{A_{\mu\nu\rho}}}{s_{ij}s_{kl}}
	\label{gluon-amp1}
\end{eqnarray}
where \begin{equation*}
	\mathcal{A_{\mu\nu\rho}}=-g_{\rho\mu}(p_{a}+p_{i}+p_{j})_{\nu}+g_{\mu\nu}(p_{i}+p_{j}+p_{k}+p_{l})_{\rho}+g_{\nu\rho}(p_{a}-p_{k}-p_{l})_{\mu}\nonumber
\end{equation*}
The last term in $\mathcal{M}_g$ can be simplified as:
\begin{equation*}
	\frac{g^{3}\,[j\,\gamma^{\mu}\,i\rangle\,[k\,\gamma^{\nu}\,l\rangle\,\varepsilon_{+}^{\rho}\,f^{abc}T^{b}_{ji}T^{c}_{kl}\,\mathcal{A_{\mu\nu\rho}}}{s_{ij}s_{kl}}
	=\frac{-\sqrt{2}\,ig^{3}\mathcal{B}}{s_{ij}s_{kl}\langle qa\rangle}(T^b _{ji} T^b _{kn} T^a _{nl} - T^b _{ji} T^a _{kn} T^b _{nl}),\nonumber
\end{equation*}
where
\begin{equation*}
	\mathcal{B}=[ja] \langle qi \rangle [ka] \langle al \rangle -[jk] \langle li \rangle \{ [ak] \langle kq \rangle + [al] \langle lq \rangle \} + [ka] \langle ql \rangle \{ [jk] \langle ki \rangle + [jl] \langle li \rangle \} \nonumber
\end{equation*}
In the next section, we calculate amplitude square $\mathcal{M}^2$ taking into account these contributions from photon-mediated diagrams and gluon-mediated diagrams. 

\section{$\mathcal{M}^{2}$ for $q\overline{q}\rightarrow q\overline{q}g$}
\label{calc_check}
\begin{figure}[t!]
	\centering
	{\includegraphics[width = 1.75in]{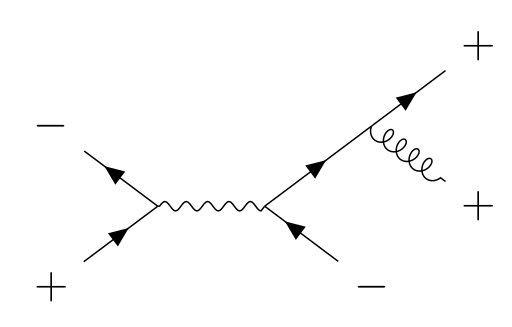}} 
	\hspace{0.6cm}
	{\includegraphics[width = 1.75in]{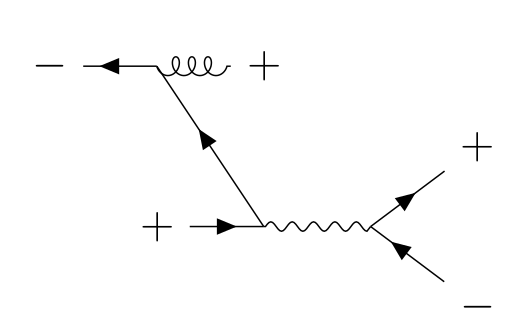}}
	\hspace{0.6cm}
	{\includegraphics[width = 1.75in]{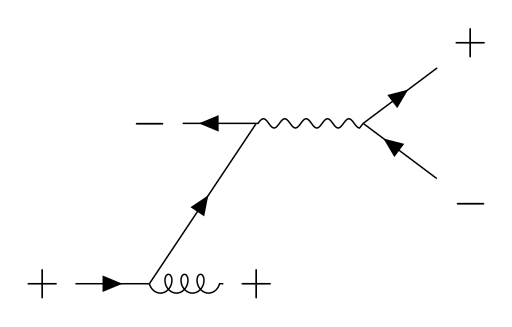}}
	\hspace{0.6cm}
	{\includegraphics[width = 1.75in]{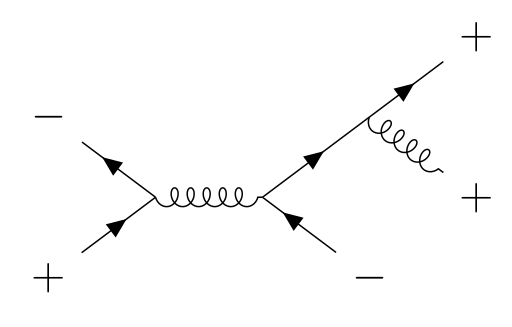}}
	\hspace{0.6cm}
	{\includegraphics[width = 1.75in]{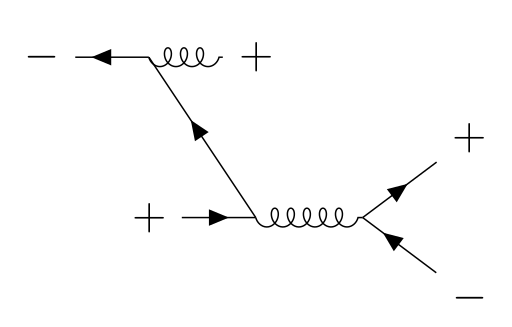}}
	\hspace{0.6cm}
	{\includegraphics[width = 1.75in]{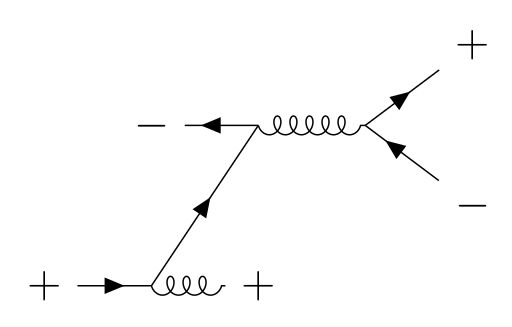}}
	\hspace{0.6cm}
	{\includegraphics[width = 1.75in]{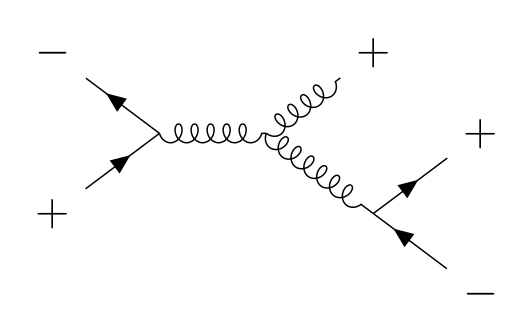}}
	\caption{Non-zero diagrams after the choice $q=l$ in $q\overline{q}\rightarrow q\overline{q}g$ process }
	\label{fig:remaining-diags}
\end{figure}

To calculate amplitude square for the process $q\overline{q}\rightarrow q\overline{q}g$, we consider Eq.~\ref{photon-amp} and Eq.~\ref{gluon-amp1}. Adding photon and gluon mediated contributions, squaring and calculating diagonal as well as interference terms using these multiple amplitude terms is complicated. Hence, to simplify the calculation, we choose the reference momentum  $q=l$\cite{Brecht} and use momentum conservation $i+j=a+k+l$. After implementing these, few amplitude terms cancel out. The non-zero amplitude terms with the corresponding diagrams (as shown in Fig~\ref{fig:remaining-diags}) add up to 

\begin{eqnarray}
	\mathcal{M}_{\gamma,g}&=& \mathcal{M}_{1} +\mathcal{M}_{2}+\mathcal{M}_{3}+\mathcal{M}_{4}+\mathcal{M}_{5}+\mathcal{M}_{6}+\mathcal{M}_{7}  \nonumber \\
	&=&\frac{\sqrt{2}\,iQ_{i}Q_{k}e^{2}g\,\langle li \rangle [ka] [j\,\slashed{i_{}}\,l\rangle \,T^{a}_{kl}\delta_{ij}}{s_{ij}s_{ak}\langle la\rangle } + \frac{-\sqrt{2}\,iQ_{i}Q_{k}e^{2}g\,\langle li\rangle [ja] [k\,\slashed{i_{}}\,l\rangle T^{a}_{ji}\delta_{kl}}{s_{kl}s_{aj}\langle la\rangle}  \nonumber\\
	&~& \frac{-\sqrt{2}\,iQ_{i}Q_{k}e^{2}g\,\langle li\rangle [kj] [a\slashed{i_{}}\,l\rangle T^{a}_{ji}\delta_{kl}}{s_{kl}s_{ai}\langle la\rangle} + \frac{\sqrt{2}\,ig^{3}\,\langle li \rangle [ka] [j\,\slashed{i_{}}\,l\rangle}{s_{ij}s_{ak}\langle la\rangle }\, T^{b}_{ji}T^{b}_{nl}T^{a}_{kn} \nonumber\\
	&~& + \frac{-\sqrt{2}\,ig^{3}\,\langle li\rangle [ja] [k\,\slashed{i_{}}\,l\rangle}{s_{kl}s_{aj}\langle la\rangle }\, T^{b}_{ni}T^{b}_{kl}T^{a}_{jn} + \frac{-\sqrt{2}\,ig^{3}\,\langle li\rangle [kj] [a\slashed{i_{}}\,l\rangle}{s_{kl}s_{ai}\langle la\rangle }\, T^{b}_{jn}T^{b}_{kl}T^{a}_{ni}\nonumber\\
	&~& \frac{-\sqrt{2}\,ig^{3} \langle li \rangle [ka][j\,\slashed{i_{}}\,l \rangle }{s_{ij}s_{kl}\langle la\rangle}(T^b _{ji} T^b _{kn} T^a _{nl} - T^b _{ji} T^a _{kn} T^b _{nl})
	\label{rem-amp}
\end{eqnarray}

The amplitude square for the process $q\overline{q}\rightarrow q\overline{q}g$ can be obtained by squaring eq.~\ref{amp}  
\begin{equation}
    |\mathcal{M}_{\gamma,g}|^2=\frac{1}{9}\left[ \sum_{i=1}^7 |\mathcal{M}_{ii}|^2 +\sum_{i<j}2\cdot Re|\mathcal{M}_{ij}|^2\right]
\end{equation}
where \(\mathcal{M}_{ii}=\mathcal{M}_i\mathcal{M}_i^*\) represent the diagonal terms  and \(\mathcal{M}_{ij}=\mathcal{M}_i\mathcal{M}_j^*\) represent the interference terms between photon mediated and gluon meditaed diagrams. The factor $1/9$ is due to averaging over initial colours. Combining diagonal and cross terms for all the diagrams, we obtain $|\mathcal{M}(i^{+}j^{-}k^{+}l^{-}a^{+})|^2$ as

\begin{eqnarray}
	|\mathcal{M}|^{2}  
   &=& 12\mathcal{A}_{11} +12\mathcal{A}_{22}+12\mathcal{A}_{33}+\frac{8}{3}\mathcal{A}_{44}+\frac{8}{3}\mathcal{A}_{55}+\frac{8}{3}\mathcal{A}_{66}+6\mathcal{A}_{77} +2\times\left(\frac{}{}2 \mathcal{A}_{15}+ 2 \mathcal{A}_{16}   + 12 \mathcal{A}_{23}  \right.\nonumber\\
	&~& \left.+  2 \mathcal{A}_{24}  +  2 \mathcal{A}_{34} - \frac{2}{3} \mathcal{A}_{45}+  \frac{7}{3}\mathcal{A}_{46}- 3 \mathcal{A}_{47}-  \frac{1}{3}  \mathcal{A}_{56} +  3 \mathcal{A}_{57}  
  - 3  \mathcal{A}_{67}\right) \nonumber \\
  \end{eqnarray}
  where $\mathcal{A}_{ij}$ with $s_{ij} = \left(p_{i} + p_{j}\right)^{2}$ are given in Table~\ref{tab:amp}
\\\\
\begin{table}[]
    \centering
\begin{tabular}{|c|c|}
	\hline 
	$\mathcal{A}_{11}=2Q_i^2Q_k^2e^4g^2\frac{s_{li}s_{ka}}{s_{ij}^2s_{aj}^2s_{la}}\textnormal{tr}\,(\slashed{i}\slashed{l}\slashed{i}\slashed{j}P_-) $ & $\mathcal{A}_{22}=2Q_i^2Q_k^2e^4g^2\frac{s_{li}s_{ja}}{s_{kl}^2s_{aj}^2s_{la}}\textnormal{tr}\,(\slashed{i}\slashed{l}\slashed{i}\slashed{k}P_-)$ \\
	\hline 
	$\mathcal{A}_{33}=2Q_i^2Q_k^2e^4g^2\frac{s_{li}s_{kj}}{s_{ij}^2s_{aj}^2s_{la}}\textnormal{tr}\,(\slashed{i}\slashed{l}\slashed{i}\slashed{a}P_-)$ & $\mathcal{A}_{44}=2g^6\frac{s_{li}s_{ka}}{s_{ij}^2s_{aj}^2s_{la}}\textnormal{tr}\,(\slashed{i}\slashed{l}\slashed{i}\slashed{j}P_-)$ \\
	\hline
	$\mathcal{A}_{55}=2g^6\frac{s_{li}s_{ja}}{s_{kl}^2s_{aj}^2s_{la}}\textnormal{tr}\,(\slashed{i}\slashed{l}\slashed{i}\slashed{k}P_-)$ & $\mathcal{A}_{66}=2g^6\frac{s_{li}s_{kj}}{s_{ij}^2s_{aj}^2s_{la}}\textnormal{tr}\,(\slashed{i}\slashed{l}\slashed{i}\slashed{a}P_-)$ \\
	\hline	
	$\mathcal{A}_{77}=2g^6\frac{s_{li}s_{ka}}{s_{ij}^2s_{aj}^2s_{la}}\textnormal{tr}\,(\slashed{i}\slashed{l}\slashed{i}\slashed{j}P_-)$ & $\mathcal{A}_{15} =-2Q_iQ_ke^2g^4\frac{s_{li}}{s_{ij}s_{kl}s_{aj}s_{ak}s_{la}}\textnormal{tr}\,(\slashed{i}\slashed{l}\slashed{i}\slashed{k}\slashed{a}\slashed{j}P_-) $   \\
	& $=\mathcal{A}_{24}$\\
	\hline
	$\mathcal{A}_{16}=-2Q_iQ_ke^2g^4\frac{s_{li}}{s_{ij}s_{kl}s_{ai}s_{ak}s_{la}}\textnormal{tr}\,(\slashed{i}\slashed{l}\slashed{i}\slashed{a}\slashed{k}\slashed{j}P_-)$  & $\mathcal{A}_{23}= -2Q_i^2Q_k^2e^4g^2  \frac{s_{li}}{s_{kl}^{2} s_{aj}s_{ai}s_{la}}\textnormal{tr}\,(\slashed{i}\slashed{l}\slashed{i}\slashed{a}\slashed{j}\slashed{k}P_-)$ \\
	$= \mathcal{A}_{43} $ & \\
	\hline
	$\mathcal{A}_{45}=-2g^6\frac{s_{li}}{s_{ij}s_{kl}s_{aj}s_{ak}s_{la}}\textnormal{tr}\,(\slashed{i}\slashed{l}\slashed{i}\slashed{k}\slashed{a}\slashed{j}P_-)$ & $\mathcal{A}_{46}=-2g^6\frac{s_{li}}{s_{ij}s_{kl}s_{ai}s_{ak}s_{la}}\textnormal{tr}\,(\slashed{i}\slashed{l}\slashed{i}\slashed{a}\slashed{k}\slashed{j}P_-)$ \\
    \hline
	$\mathcal{A}_{47}=-2g^6\frac{s_{li}}{s_{ij}^2s_{ak}s_{kl}s_{la}}\textnormal{tr}\,(\slashed{i}\slashed{l}\slashed{i}\slashed{j}P_{-})$ & $\mathcal{A}_{56}=-2g^6\frac{s_{li}}{s_{kl}^2s_{aj}s_{ai}s_{la}}\textnormal{tr}\,(\slashed{i}\slashed{l}\slashed{i}\slashed{a}\slashed{j}\slashed{k}P_-)$ \\
	\hline	
	$\mathcal{A}_{57}=2g^6\frac{s_{li}}{s_{ij}s_{aj}s_{kl}^2s_{la}}\textnormal{tr}\,(\slashed{i}\slashed{l}\slashed{i}\slashed{j}\slashed{a}\slashed{k}P_{-})$ & $\mathcal{A}_{67}=2g^6\frac{s_{li}}{s_{ij}s_{ai}s_{kl}^2s_{la}}\textnormal{tr}\,(\slashed{i}\slashed{l}\slashed{i}\slashed{j}\slashed{k}\slashed{a}P_{-})$ \\ 
	\hline	
\end{tabular}
 \caption{ Non-zero contributions of $\mathcal{A}_{ij}$}
    \label{tab:amp}
\end{table}
$$ $$
Incorporating traces in $\mathcal{A}_{ij}$ calculated using FeynCalc, we get the amplitude square as
  \begin{eqnarray}
 |\mathcal{M}|^{2} &=& \frac{8 g^2 }{3 s_{ai} s_{aj}^2 s_{ak} s_{al}
   s_{ij}^2 s_{kl}^2 s_{la}} \nonumber \\
   &~&    \left(72 Q_i^2 Q_k^2 e^4 a\cdot k~ i\cdot j~
  i\cdot l~  s_{aj} s_{ak} s_{al} s_{ij}^2
   -72 Q_i^2 Q_k^2 e^4 a\cdot j ~i\cdot k~ i\cdot
   l ~ s_{aj} s_{ak} s_{al} s_{ij}^2 \right.\nonumber \\
   &~&   \left.  -72 Q_i^2 Q_k^2 e^4 a\cdot i~   i\cdot l~ j\cdot k~  s_{aj}
   s_{ak} s_{al} s_{ij}^2  +36 Q_i^2
   Q_k^2 e^4 i a\cdot j ~ k\cdot l~  s_{aj} s_{ak} s_{al} s_{ij}^2 \right.\nonumber \\
   &~&    \left. +36 Q_i^2 Q_k^2 e^4
  a\cdot i~ i\cdot l~  s_{ai}   s_{ak} s_{al} s_{ja} s_{kl}^2  +36 Q_i^2 Q_k^2 e^4 i\cdot j~ i\cdot l~ s_{ai} s_{ak} s_{ka} s_{kl}^2
   s_{la} \right.\nonumber \\
   &~&    \left. -18 Q_i^2 Q_k^2 e^4 k\cdot l~  s_{ai}
   s_{ak} s_{ij}^2 s_{ja} s_{la} -24 g^2 Q_i Q_k e^2 a\cdot k~
   i\cdot j~ i\cdot ~  s_{aj}^2
   s_{al} s_{ij} s_{kl} \right.\nonumber \\
   &~&    \left.  +24 g^2 Q_i Q_k e^2 a\cdot j~
   i\cdot k~ i\cdot l~  s_{aj}^2 s_{al} s_{ij} s_{kl} -24
   g^2 Q_i Q_k e^2 a\cdot i~ i\cdot l~ j\cdot
   k~  s_{aj}^2 s_{al} s_{ij} s_{kl} \right.\nonumber \\
   &~&    \left.-24 g^2  Q_i Q_k e^2 a\cdot k~ i\cdot j~
   i\cdot l~ s_{ai} s_{aj} s_{al} s_{ij}
   s_{kl}  -24 g^2 Q_i Q_k e^2 a\cdot j~ i\cdot k
   i\cdot l  s_{ai} s_{aj} s_{al} s_{ij}
   s_{kl} \right.\nonumber \\
   &~&    \left. +24 g^2 Q_i Q_k e^2 a\cdot i~ i\cdot l~ j\cdot k~  s_{ai} s_{aj} s_{al} s_{ij} s_{kl}  -2 g^4 a\cdot k
   i\cdot j~ i\cdot l~ s_{aj} s_{ak} s_{al} s_{ij}^2 \right.\nonumber \\
   &~&   \left.   +2 g^4 a\cdot j~ i\cdot k~ i\cdot l~ s_{aj} s_{ak} s_{al}  s_{ij}^2 +2 g^4 a\cdot i~ i\cdot l~ j\cdot k~
   s_{aj} s_{ak} s_{al} s_{ij}^2 +26 g^4 i\cdot j~ i\cdot l s_{ai} s_{ak} s_{al}  s_{ka} s_{kl}^2 \right.\nonumber \\
   &~&   \left. +8 g^4 a\cdot i~ i\cdot l~ s_{ai} s_{ak} s_{al}
   s_{kj} s_{kl}^2  -18 g^4 a\cdot k i\cdot j~ i\cdot l~ s_{aj}^2 s_{ak} s_{al}
   s_{ij} \right.\nonumber \\
   &~&   \left. +18 g^4 a\cdot j~ i\cdot k~ i\cdot l~ s_{aj}^2 s_{ak} s_{al} s_{ij}-18 g^4 a\cdot i~
   i\cdot l~ j\cdot k~ s_{aj}^2 s_{ak} s_{al} s_{ij}  \right.\nonumber \\
   &~&   \left. +18 g^4 a\cdot k~ i\cdot j~ i\cdot l~ s_{ai} s_{aj} s_{ak} s_{al} s_{ij} +18 g^4 a\cdot j~ i\cdot  k~ i\cdot l~ s_{ai} s_{aj} s_{ak}
   s_{al} s_{ij}\right.\nonumber \\
   &~&    \left.-18 g^4 a\cdot i~ i\cdot l~ j\cdot k~ s_{ai} s_{aj} s_{ak} s_{al} s_{ij}+8 g^4 i\cdot k i\cdot l~
   s_{ai} s_{ak} s_{al} s_{ij}^2 s_{ja} +18 g^4
   i\cdot j~ i\cdot l~ s_{ai} s_{aj}^2
   s_{al} s_{kl} \right.\nonumber \\
   &~&    \left. +14 g^4 a\cdot j~ i\cdot k~   i\cdot l~ s_{aj}^2 s_{al} s_{ij} s_{kl} -14 g^4 a\cdot i~   i\cdot l~ j\cdot k~ s_{aj}^2 s_{al}   s_{ij} s_{kl}\right.\nonumber \\
   &~&    \left. +4 g^4 a\cdot k~ i\cdot j~ i\cdot l~ s_{ai} s_{aj} s_{al}
   s_{ij} s_{kl}+4 g^4 a\cdot j~ i\cdot   k~ i\cdot l~ s_{ai} s_{aj} s_{al}
   s_{ij} s_{kl}    
   \right.\nonumber \\
   &~&    \left. -4 g^4   a\cdot i~ i\cdot l~ j\cdot k~ s_{ai} s_{aj} s_{al} s_{ij} s_{kl} +s_{ai} \left(9 g^4 s_{ak}-2 g^2 \left(g^2-6 e^2 t
   Q_i Q_k\right) s_{kl}\right) \right.\nonumber \\
   &~&    \left.+s_{aj} \left(9 g^4 s_{ak}-g^2 \left(12  Q_i Q_k e^2+7 g^2\right)
   s_{kl}\right)\right) s_{li}
	\label{rem-amp}
\end{eqnarray}

\section{Conclusion}
In this work, we have calculated $S$ matrix elements for the process $q\overline{q}\rightarrow q\overline{q}g$ mediated by photon and gluon at the tree level using spinor-helicity formalism. Out of four photon exchange diagrams and five gluon exchange diagrams in spinor-helicity formalism, we get seven non zero terms in $\mathcal{M}^{2}$ as two diagonal terms and many interference terms in S matrix elements vanish. Choice of suitable reference momentum simplifies the calculation to a great
extent. Many tree level as well as loop level calculations can be simplified
using this technique. Hence, spinor-helicity formalism can be used as as an efficient method to
simplify the complexity of presicion calculations.

We note that in the amplitude square, contributions due to the photon exchange are $e^4g^2$ in cross-term whereas contribution due to the gluon exchange are  $g^6$ in cross-term. For the photon-gluon interference terms, we have $e^2g^4$. From this, one can qualitatively understand the variation in $\alpha_s$ contribution for photon and gluon in the $q\overline{q}\rightarrow q\overline{q}g$ scattering process mediated by photon and gluon. This process can also be mediated by $Z$ boson which will have vector and axial vector structure in couplings. The electroweak mediators will contribute to same order in $\alpha_s$ as the photon. It will be interesting to study how the results change when we include all - QCD, QED and electroweak mediators.  

\section*{Acknowledgements}
VP gratefully acknowledges the support of the Department of Atomic Energy, Government of India, under Project Identification No. RTI4001. This work was carried out as part of the National Initiative on Undergraduate Science (NIUS) program of the Homi Bhabha Centre for Science Education, Tata Institute of Fundamental Research (HBCSE-TIFR), Mumbai, India. AM would like to thank Department of Atomic Energy for the award of Raja Ramanna Chair.
\newpage
\appendix
\section{Feynman Rules}
\renewcommand{\theequation}{A.\arabic{equation}}
\setcounter{equation}{0}
\vspace{1cm}
\begin{itemize}
\vspace{-1.0cm}
\item Propagator for internal lines:
\item[] Massless vector boson: \\ \\
\begin{figure}[h!] 
\vspace*{-0.45in} 
\centerline{\includegraphics[width=0.2 \textwidth]{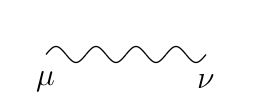} } \hspace*{-0.2in} 
\end{figure} \vspace*{-2cm}
$\frac{-ig_{\mu\nu}}{p^2+i\epsilon}$
\vspace{1cm}

\item[]
\item Vertex factor:
\item[] QED vertex = $-i e \gamma_\mu$ \hspace{6.0cm}
\item[] QCD vertex = $-i g \gamma_\mu T^a$
\begin{figure}[h!] 
\vspace*{-2.6cm} 
\centerline{\includegraphics[width=0.25 \textwidth]{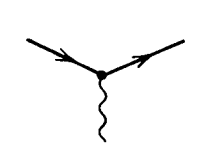} } \hspace*{-0.2in} 
\end{figure} 
\item[] where $T^a$ are the colour factors
\item[]
\end{itemize}

\vspace{-1.0cm}
\section{Spinor-Helicity Formalism}
For a massless fermion, the 4-component spinor representation \(u(k_{i})\) of external particles can be decomposed into two 2-spinors. These spinors are eigenstates of the helicity operator with eigenvalues \(\pm \frac{1}{2}\) (right- and left-handed respectively). Since they're also eigenstates of \(\gamma^{5}=i\gamma^{0}\gamma^{1}\gamma^{2}\gamma^{3}\) with eigenvalues \(\pm 1\). Therefore, the operators \(\frac{1}{2}(1\pm\gamma^{5})\) project onto positive and negative helicity states respectively. Similarly, an anti-fermion is represented by \(v(k_{i})\). However, a negative-energy 2-spinor which describes a right-handed (positive helicity) antiparticle has negative chirality. For this reason, we use the same 2-spinor for a left-handed particle and a right-handed anti-particle. We use the bra-ket notation to denote a 2-spinor:
\begin{equation*}
	|i\rangle =u_{+}(k_{i})=v_{-}(k_{i}), \qquad |i]=u_{-}(k_{i})=v_{+}(k_{i}),
\end{equation*}
\begin{equation*}
	\langle i|=\overline{u_{-}}(k_{i})=\overline{v_{+}}(k_{i}),\qquad [i|=\overline{u_{+}}(k_{i})=\overline{v_{-}}(k_{i}).
\end{equation*}
These 2-spinors satisfy the following inner-product identities:
\begin{equation*}
	\langle i|j\rangle =-\langle j|i\rangle ,\quad [i|j]=-[j|i],\quad \langle i|j]=0=[i|j\rangle , \quad \langle i|j\rangle ^{*} = [j|i]
\end{equation*}
Using the (massless) completeness relation for \(u_{s}(k)\),
\begin{equation*}
	\sum_{s}u_{s}(k)\overline{u_{s}(k)}=|k\rangle[k|+|k]\langle k|=\slashed{k}.
\end{equation*}
The following identities are particularly useful:
\subsubsection*{Gordon Identity:}
\begin{equation*}
	[i\,\gamma^{\mu}\,i\rangle =2\,k^{\mu}_{i}
	; \quad
	[ij]\langle ji\rangle=2\,k_{i}\cdot k_{j}=s_{ij}
\end{equation*}
\subsubsection*{Fierz Rearrangement:}
\begin{equation*}
	[i\,\gamma^{\mu}\,j\rangle[k\,\gamma_{\mu}\,l\rangle=2\,[ik]\langle lj\rangle\,;\quad
	\gamma_{\mu}[i\,\gamma^{\mu}\,j\rangle =|i]\langle j|+|j\rangle [i|
\end{equation*}
The object $[j|\gamma^{\mu}|i\rangle$ transforms as a 4-vector. This can be used to represent the polarization vector $\varepsilon^{\mu}$ in terms of spinor products. For a vector boson having momentum $k$, we can use an arbitrary momentum $q$ to write down polarizations of fixed helicity:
\begin{equation*}
	\varepsilon^{\mu}_{+}(k,q)=\frac{[ k\,\gamma^{\mu}\,q\rangle}{\sqrt{2}\langle qk\rangle}, \quad
	\varepsilon^{\mu}_{-}(k,q)=\frac{[ q\,\gamma^{\mu}\,k\rangle}{\sqrt{2}[kq]}
\end{equation*}
The slashed matrices of $\varepsilon_{\pm}$ are then given by
\begin{equation*}
	\slashed{\varepsilon}_{+}(k,q)=\frac{|q\rangle[k|\,+\,|k]\langle q|}{\sqrt{2}\langle qk\rangle}, \quad
	\slashed{\varepsilon}_{-}(k,q)=\frac{|q]\langle k|\,+\,|k\rangle [q|}{\sqrt{2}[kq]}
\end{equation*}
Here $q$ is arbitrary, except $q\cdot k \neq 0$. A clever choice of $q$ may help simplify the amplitude expression. The final result is independent of the choice of $q$.

\subsection*{Colour Factors}
The generators of SU(3) satisfy
\begin{equation*}
	[T^{a},T^{b}]=if^{abc}T^{c}
\end{equation*}
\begin{equation*}
	\textnormal{tr}T^{a}T^{b}=\frac{1}{2}\delta^{ab}
	\implies \textnormal{tr}T^{a}T^{a}=T^{a}_{ij}T^{a}_{ji}=4
\end{equation*}
\[T^aT^a=\frac{4}{3}I\]
\begin{equation*}
	T^{a}_{ij}\,T^{a}_{kl}=\frac{1}{2}(\delta_{il}\delta_{jk}-\frac{1}{N}\delta_{ij}\delta_{kl})
\end{equation*}

\[\textnormal{tr}[T^aAT^aB]=\frac{1}{2}(\textnormal{tr}[A]\textnormal{tr}[B]-\frac{1}{3}\textnormal{tr}[AB])\]

\[\textnormal{tr}[T^aA]\textnormal{tr}[T^aB]=\frac{1}{2}(\textnormal{tr}[AB]-\frac{1}{3}\textnormal{tr}[A]\textnormal{tr}[B])\]

\bibliographystyle{JHEP}
\bibliography{draft}

\end{document}